# Silver Staining of 2D Electrophoresis Gels

Cécile Lelong, Mireille Chevallet, Sylvie Luche, Thierry Rabilloud

CEA-DSV-iRTSV/LBBSI and UMR CNRS 5092
CEA Grenoble
17 rue des martyrs, F-38054 Grenoble Cedex 9, France

1. Introduction

Silver staining of polyacrylamide gels was introduced in 1979 by Switzer et al. [1], and rapidly gained popularity owing to its high sensitivity, ca. 100 times higher than staining with Coomassie Blue. However, the first silver staining protocols were not trouble-free. High backgrounds and silver mirrors were frequently experienced, with a subsequent decrease in sensitivity and reproducibility. This led many authors to suggest improved protocols, so that more than 100 different silver staining protocols for proteins in polyacrylamide gels can be found in the literature. However, all of them are based on the same principle (see [2] and [3] for details) and comprise more or less four major steps.

a)  The first step is fixation, and aims at insolubilizing the proteins in the gels and removing the interfering compounds present in the 2D gels (glycine, Tris, SDS and carrier ampholytes present at the bottom of the gels).

b)  The second step is sensitization, and aims at increasing the subsequent image formation. Numerous compounds have been proposed for this purpose. all these compounds bind to the proteins, and are also able either to bind silver ion, or to reduce silver ion into metallic silver, or to produce silver sulfide [2], [3]. this sensitization step is sometimes coupled with the fixation step.

c)  The third step is silver impregnation. Either plain silver nitrate or ammoniacal silver can be used.

d)  The fourth last step is image development. For gels soaked with silver nitrate, the developer contains formaldehyde, carbonate and thiosulfate. The use of the latter compound, introduced by Blum et al [4], reduces dramatically the background and allows for thorough development of the image. For gels soaked with ammoniacal silver, the developer contains formaldehyde and citric acid. In this case, thiosulfate is better introduced at the gel polymerization step [5]. Background reduction by thiosulfate can also be achieved by brief incubation in thiosulfate prior to development [6], or by inclusion in the developer .

When the desired image level is obtained, development is stopped by dipping the gel in a stop solution,  generally containing acetic acid and an amine to reach a pH of 7. Final stabilization of the image is achived by thorough rinsing in water to remove all the compounds present in the gel. However, the development of downstream  protein characterisation methods, such as analysis by mass spectrometry starting from gel separated proteins, has brought additional constraints on silver staining. Besides the classical constraints weighing on any detection method (sensitivity, linearity and homogeneity), the interface of the detection method with downstream methods becomes more and more important. In the case of mass spectrometry, this interface comprises the compatibility with enzymatic digestion and peptide extraction, as well as the absence of staining-induced peptide modifications. While the exquisite sensitivity of silver staining is unanimously recognized, its compatibility with downstream analysis appears more problematic than staining with organic dyes (e.g. Coomassie Blue). A mechanistic study [7] has shown that these problems are linked in part to the pellicle of metallic silver deposited on the proteins during staining, but are mainly due to the presence of formaldehyde during silver staining. Up to now, formaldehyde is the only chemical known able to produce a silver image of good quality in protein staining, and attempts to use other chemicals have proven rather unsuccessful [7]. However, besides a lowered peptide representation in silver stained gels [7], formaldehyde induces some peptide modifications, such as +12  and +30 Da adducts [8] as well as formylation [9], the latter being most likely caused by the formic acid produced upon reaction of formaldehyde with silver ions in the image development step.

These problems are common to all silver staining protocols, although their extent is variable from one protocol to another. Some guidelines for the choice of a silver staining protocol are described in Note 1

2. Materials

2.1. Equipment

1. Glass dishes or polyethylene food dishes. The latter are less expensive, have a cover and can be easily piled up for multiple staining. They are however more difficult to clean, and it is quite important to avoid scratching of the surface, which will induce automatic silver deposition in subsequent stainings. Traces of silver are generally easily removed by wiping the plastic box with a tissue soaked with ethanol. If this treatment is not sufficient, stains are easily removed with Farmer's reducer (0.1% sodium carbonate, 0.3% potassium hexacyanoferrate (III) and 0.6% sodium thiosulfate). Thorough rinsing of the box with water and ethanol terminate the cleaning process.
2. Plastic sheets (e.g. the thin polycarbonate sheets sold by Bio-Rad for multiple gel casting) Used for batch processing.
3. Reciprocal Shaking platform: The use of orbital or three-dimensional movement shakers is not recommended.

2.2. Reagents

Generally speaking, chemicals are of standard pro analysis grade.

1. Water: The quality of the water is of great importance. Water purified by ion exchange cartridges, with a resistivity greater than 15 MegaOhms/ cm, is very adequate, while distilled water gives more erratic results.
2. Formaldehyde: Formaldehyde stands for commercial 37-40% formaldehyde. This is stable for months at room temperature. It should not be stored at 4°C, as this promotes polymerization and deposition of formaldehyde. The bottle should be discarded when a layer of polymer is visible at the bottom of the bottle.
3. Sodium thiosulfate solution: 10% (w/v) solution of crystalline sodium thiosulfate pentahydrate in water. Small volumes of this solution (e.g. 10 ml) are prepared fresh every week and stored at room temperature.
4. Ethanol: A technical grade of alcohol can be used, and 95% ethanol can be used instead of absolute ethanol, without any volume correction. The use of denatured alcohol is however not recommended.
5. Citric acid solution: 2 M citric acid. Stored at room temperature for months.
6. Potassium acetate solution: 5 M potassium acetate. This is stable for months at room

temperature.

7. Silver nitrate solution: 1 N silver nitrate. A 1 N silver nitrate solution (Fluka) is less expensive than solid silver nitrate, and is stable for months if kept in a fridge (a black and cold place).
8. Sodium hydroxide solution: 1 N sodium hydroxide (Fluka).
9. Ammonium hydroxide solution: 5 N ammonium hydroxide (Aldrich). Ammonium hydroxide is kept in the fridge.

2.2.1. Solutions for Protocol 1
1. Fix solution I: 25% v/v ethanol, 4% v/v formaldehyde.
2. Sensitivity enhancing solution I: 0.05% (w/v) 2-7 Naphtalene disulfonate disodium salt (NDS)(Acros chemicals)
3. Ammoniacal silver solution: To prepare 500 ml of this solution,(sufficient for a batch of 4 gels) place 480 ml of water in a flask under strong magnetic agitation. Add successively 7.5 ml of 1N sodium hydroxide, 7.5 ml of 5N ammonium hydroxide and 12 ml of 1N silver nitrate. Upon addition of silver nitrate, a transient brown precipitate forms, which should redissolve within a few seconds (see Note 3). Only clear solutions should be used.
4. Development solution I: 1 ml of 37% formaldehyde and 180µl of 2M citric acid per liter (see Note 4).
5. Stop solution I: 20 ml acetic acid and 5 ml ethanolamine per liter.

2.2.2. Solutions for Protocol 2
1. Fix solution II: 5% acetic acid/30% ethanol and 0.05% (w/v) 2-7 Naphtalene disulfonate disodium salt (NDS)(Acros chemicals).
2. Ammoniacal silver solution: To prepare 500 ml of this solution,(sufficient for a batch of 4 gels) place 480 ml of water in a flask under strong magnetic agitation. Add successively 7.5 ml of 1N sodium hydroxide, 7.5 ml of 5N ammonium hydroxide and 12 ml of 1N silver nitrate. Upon addition of silver nitrate, a transient brown precipitate forms, which should redissolve within a few seconds (see Note 3). Only clear solutions should be used.
3. Development solution I: 1 ml of 37% formaldehyde and 180µl of 2M citric acid per liter (see Note 4).

4. Stop solution I: 20 ml acetic acid and 5 ml ethanolamine per liter.

2.2.3. Solutions for Protocol 3

1. Fix solution III: 5% acetic acid, 30% ethanol.
2. Sensitivity enhancing solution II: 2ml of 10% thiosulfate per liter.
3. Silver stain solution II: and 12.5 ml of 1 N silver nitrate per liter.
4. Development solution II: 30 g anhydrous potassium carbonate, 250µl 37% formaldehyde and 125µl 10% thiosulfate per liter (see Note 9).
5. Stop solution II: 40g of Tris and 20 ml of acetic acid per liter.

2.2.4. Solutions for Protocol 4

1. Fix solution IV: 5% acetic acid/30% ethanol.
2. Sensitivity enhancing solution III: 0.5 M potassium acetate, 25% ethanol and 3 g potassium tetrathionate per liter (see Note 10).
3. Silver stain solution III: 12.5ml 1N silver nitrate per liter.
4. Development solution III: 30 g anhydrous potassium carbonate, 250µl 37% formaldehyde and 125µl 10% thiosulfate per liter (see Note 9).
5. Stop solution III: 40g of Tris and 20 ml of acetic acid per liter.

3. Methods

3.1. General practice

Batches of gels (up to 4 gels per box) can be stained. For a batch of 3 or 4 medium-sized gels (e.g. 160x200x1.5 mm), 1 liter of the required solution is used, which corresponds to a solution/gel volume ratio of at least 5. 500 ml of solution is used for 1 or 2 gels. Batch processing can be used for every step longer than 5 minutes, except for image development, where one gel per box is required. For steps shorter than 5 minutes, the gels should be dipped individually in the corresponding reagent(s).

For changing solutions, the best way is to use a plastic sheet . This is pressed on the pile of gels with the aid of a gloved hand. Inclining the entire setup allows to empty the box while keeping the gels in it. The next solution is poured with the plastic sheet in place, which prevents the flow to fold or break the gels. The plastic sheet is removed after the solution change and kept in a separate box filled with water until the next solution change. This water is changed after each complete round of silver staining.

When gels must be handled individually, the are manipulated with gloved hands. The use of powder-free, nitrile, gloves is strongly recommended, as standard gloves are often the cause of pressure marks.

Except for development or short steps, where occasional hand agitation of the staining vessel is convenient, constant agitation is required for all the steps. A reciprocal ("ping-pong") shaker is used at 30-40 strokes per minutes.

Four different silver staining protocols are detailed below. The rationale for choosing one of them according to the constraints brought by the precise 2D protocol used and the requisites of the experimentator are described in Note1.

3.2. PROTOCOL 1: silver staining with ammoniacal silver and formaldehyde fixation

This protocol is based on the original protocol of Eschenbruch and Bürk [10], with modifications [5, 11,12]. For optimal results, alterations must be brought at the level of gel casting. Piperazine diacrylamide is used as crosslinker in place of Bis (in a weight to weight substitution) [11], but this is not mandatory, and thiosulfate is added at the gel polymerization step [5]. Practically, the initiating system is composed of 1µl of TEMED, 10µl of 10% sodium thiosulfate solution and 10µl of 10% ammonium persulfate solution per ml of gel mix. This ensures correct gel formation and gives minimal background upon staining.

After electrophoresis, silver staining processes as follows:

1. Unmold the gels in water, and let rinse for 5-10 minutes.
2. Soak gels in fix solution I for 1 hour
3. Rinse 2x 15 minutes in water
4. Sensitize overnight in sensitivity enhancing solution I.
5. Rinse 6x 20 minutes in water
6. Impregnate for 30-60 minutes in the ammoniacal silver solution.
7. Rinse 3x 5 minutes in water
8. Develop image (5-10 minutes) in development solution I.

9. Stop development in stop solution I. Leave in this solution for 30 to 60 minutes.
10. Rinse with water (several changes) prior to drying or densitometry.

3.3. PROTOCOL 2 : silver staining with ammoniacal silver [13]

1. Fix the gels in fix solution II (1 hour + overnight) (see Note 2).
3. Rinse in water (6x 15 minutes).
4. Impregnate for 30-60 minutes with ammoniacal silver solution
5. Rinse 3x 5 minutes in water
6. Develop image (5-10 minutes) in development solution I.
7. Stop development in stop solution I. Leave in this solution for 30 to 60 minutes.
8. Rinse with water (several changes) prior to drying or densitometry.

3.4. PROTOCOL 3: fast silver staining

This protocol is based on the protocol of Blum et al. [4], with modifications [12].

1. Soak the gels in fix solution II for at least 3x 30 minutes (see Note 5).
2. Rinse in water for 3x 10 minutes.
3. To sensitize, soak gels for 1 minute (1 gel at a time) in sensitivity enhancing solution II.
4. Rinse 2x 1 minute in water (see Note 6).
5. Impregnate for at least 30 minutes in silver solution II (see Note7).
6. Rinse in water for 5-15 seconds (see Note 8).
7. Develop image (10-20 minutes) in development solution II (see Note9).
8. Stop development (30-60 minutes) in stop solution II.
9. Rinse with water (several changes) prior to drying or densitometry.

3.5. PROTOCOL 4 : long silver nitrate staining for free-floating gels [14]

1. Fix the gels in fix solution IV (3x 30 minutes).
2. Sensitize overnight in sensitivity enhancement solution IV.
3. Rinse in water (6 x 20 minutes).
4. Impregnate for 1-2 hours with silver in a silver solution IV.
5. Rinse with water for 5-10 seconds (see Note 8).
6. Develop image (10-20 minutes) in development solution IV (see Note 9).

7. Stop development (30-60 minutes) in stop solution IV.
8. Rinse with water (several changes) prior to drying or densitometry.

4. Notes

1. From the rather simple theoretical bases described in the introduction, more than 100 different protocols were derived. The changes from one protocol to another are present either in the duration of the different steps, or in the composition of the solutions. The main variations concern either the concentration of the silver reagent, or the nature and concentration of the sensitizers. Only a few comparisons of silver staining protocols have been published [12, 15]. From these comparisons, selected protocols have been proposed in the former sections. The choice of a protocol will depend on the constraints of the experimental setup and of the requisites of the experimentator (speed, reproducibility, etc.). The following guidelines can be suggested.
    a) The maximum sensitivity is not widely different from one protocol to another. The main differences are in the uniformity of the staining from one protein to another, the reproducibility of the staining, the speed of the method, and its adaptation to the various 2D protocols. For example, the carrier ampholytes that are present in all 2D protocols require rigorous fixation steps which are not mandatory when simple SDS gels are to be silver stained. On a different side, there is a trade off between speed and long-term reproducibility. Fast protocols use short steps (less than 5 minutes) and thus a more transient chemistry, which are difficult to keep reproducible
    b) Generally speaking, methods using ammoniacal silver give very uniform results, with minimal color effects and improved compatibility with mass spectrometry [13]. They are by far more sensitive than silver nitrate-based methods for the staining of basic proteins, and are therefore strongly recommended for 2D gels with very wide pH gradients. However, these methods suffer from a number of minor drawbacks which prevent their universal use.
    c) The silver reagent is very sensitive to the ammonia concentration. As ammonia is highly volatile, this introduces problems for the long term reproducibility of the

method. This problem can be alleviated to a large extent by the use of commercial titrated ammonia solutions.

d) Ammoniacal silver is not compatible with all SDS gel systems. Systems using Tricine or Bicine as trailing ions are not compatible with ammoniacal silver staining.

e) Ammoniacal silver staining is not recommended for gels supported by a plastic film. Silver mirrors are frequently encountered in this case.

f) Optimal protocols for ammoniacal silver staining (e.g. Protocol 1 in section 3) are generally time-consuming. In addition, optimal results are obtained with the use of home-made gels, containing PDA as a crosslinker [11] and with thiosulfate included at the gel polymerization step [5]. This prevents the use of commercial gels. Moreover, these protocols give best results when aldehydes (formaldehyde or glutaraldehyde) are used as fixers/sensitizers. This prevents any recovery of the silver-stained protein for subsequent use (e.g. Mass spectrometry ). This drawback can be however alleviated (see protocol 2), at the expense of the uniformity of the staining (see figure 1).

g) Silver staining being a delicate process [2,3], the temperature control in the laboratory plays a role in the optimal silver staining protocol. Ammoniacal silver is hampered by low temperatures (less than 18°C) while high temperatures (more than 30°C) produce a yellow background in silver nitrate staining

2. Other fixation processes can be used. For gels running overnight, a shorter process can be used. For ammoniacal silver staining (protocol 2 ), fix the gel in fixing solution II for 3x30 minutes, then rinse in water for 4x 15 minutes. Return to standard protocol at step 4 (silver impregnation).
For silver nitrate staining, fixation can be reduced to a single 30 minutes bath [16] . This will improve sequence coverage in mass spectrometry, at the expense of a strong chromatism (spots can be yellow, orange, brown or grey), making image analysis difficult. Furthermore, ampholytes are not removed by short fixation and give a grey background at the bottom of the 2D gels

3. The composition of the ammoniacal silver solution has an important influence on the final

sensitivity. The ammonia/silver mole ratio is in fact the key parameter [10]. The solution given in this protocol has an ammonium/silver molar ratio of 3.1, which ensures maximal sensitivity, but less stability of the solution. If a brown precipitate remains in the solution, this means that the ammonia solution is no longer concentrated enough. The best remedy is to discard the ammoniacal silver solution and to prepare a new one with a new bottle of 5N ammonium hydroxide. If this is not possible, add small aliquots of ammonia to the precipitated ammoniacal silver solution until the solution becomes clear. The sensitivity will be however lesser than usual. If reduced sensitivites are required, increase the ammonium hydroxide concentration by a factor of 1.3 to 2. this will progressively decrease the sensitivity.

4. In a standard analytical 2D gel loaded with 50-100µg of protein, the first major spots should begin to appear within 1 minute. Delayed appearance indicates lower than expected sensitivity, but is observed when aldehydes have not been used in the fixing process. In the latter case, sensitivity is restored by a longer development. The developer should be altered if no thiosulfate is present in the gel (e.g. use of ready-made gels). To prevent the rapid appearance of background, add 10µl of 10% thiosulfate per liter of developer. The maximum sensitivity will not be altered by this variation. However, some spots will show a lighter color or give hollow spots.

5. The fixation process can be altered if needed. The figures indicated in the protocol are the minimum times. Gels can be fixed without any problem for longer periods. For example, gels can be fixed overnight, with only one solution change. For ultra rapid fixation, the following process can be used [16]:

    6. Fix in 10% acetic acid /40% ethanol for 10 minutes, then rinse for 10 minutes in water.

    7. Post fix in 0.05% glutaraldehyde/40% ethanol and 100µl/l 37% formaldehyde for 5 minutes

    8. Rinse in 40% ethanol for 2x 10 minutes then in water for 2x 10 minutes. Proceed to step3

6. The optimal setup for sensitization is the following. Prepare four staining boxes containing respectively the sensitizing thiosulfate solution, water (2 boxes), and the silver nitrate solution. Put the vessel containing the rinsed gels on one side of this series of boxes. Take one gel out of the vessel and dip it in the sensitizing and rinsing solutions ( 1 minute in each solution). Then transfer to silver nitrate. Repeat this process for all the gels of the batch. A new gel can be sensitized while the former one is in the first rinse solution, provided that the 1 minute time is kept (use a bench chronometer). When several batches of gels are stained on the same day , it is necessary to prepare several batches of silver solution. However, the sensitizing and rinsing solutions can be kept for at least three batches, and probably more.

7. Gels can be impregnated with silver for at least 30 minutes and at most 2 hours without any change in sensitivity or background.

8. This very short step is intended to remove the liquid film of silver solution brought with the gel.

9. When the gel is dipped in the developer, a brown microprecipitate of silver carbonate should form. This precipitate must be redissolved to prevent deposition and background formation. This is simply achieved by _immediate_ agitation of the box. Do not expect the appearance of the major spots before 3 minutes of development. The spot intensity reaches a plateau after 15-20 minutes of development, and then background appears. Stop development at the beginning of background development. This ensures maximal and reproducible sensitivity.

10. The sensitization solution is prepared as follows. Dissolve potassium tetrathionate in water (half the desired final volume). After complete dissolution, add the required volumes of concentrated potassium acetate and ethanol. Fill up to the final volume with water.

Figures

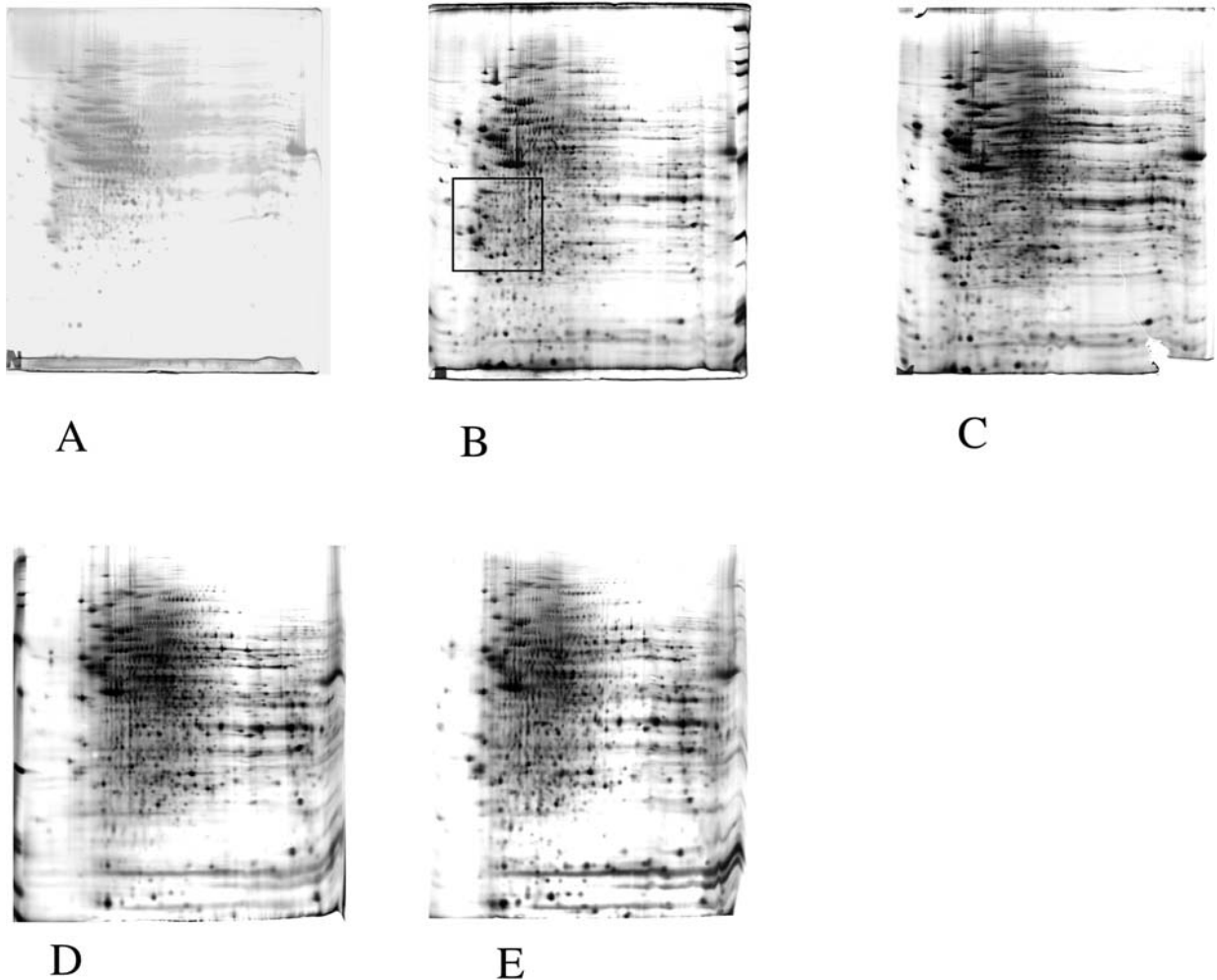

Figure1: global comparison of silver-stained gels.
100 micrograms of proteins prepared from a complete cell extract (WEHI 274 monocytic cells) are loaded on two-dimensional gels (pH 3-10.5 pH range, 10% acrylamide). The gels are then stained by different silver staining methods described in the chapter. A: ultrafast method; B: fast silver nitrate; C: long silver nitrate; D: Ammoniacal silver without aldehyde fixation; E: Ammoniacal silver with formaldehyde fixation. The Ammonical silver methods are more sensitive with basic proteins but less with acidic proteins, as shown in Figure 2. Box: gel zone magnified in figure 2.

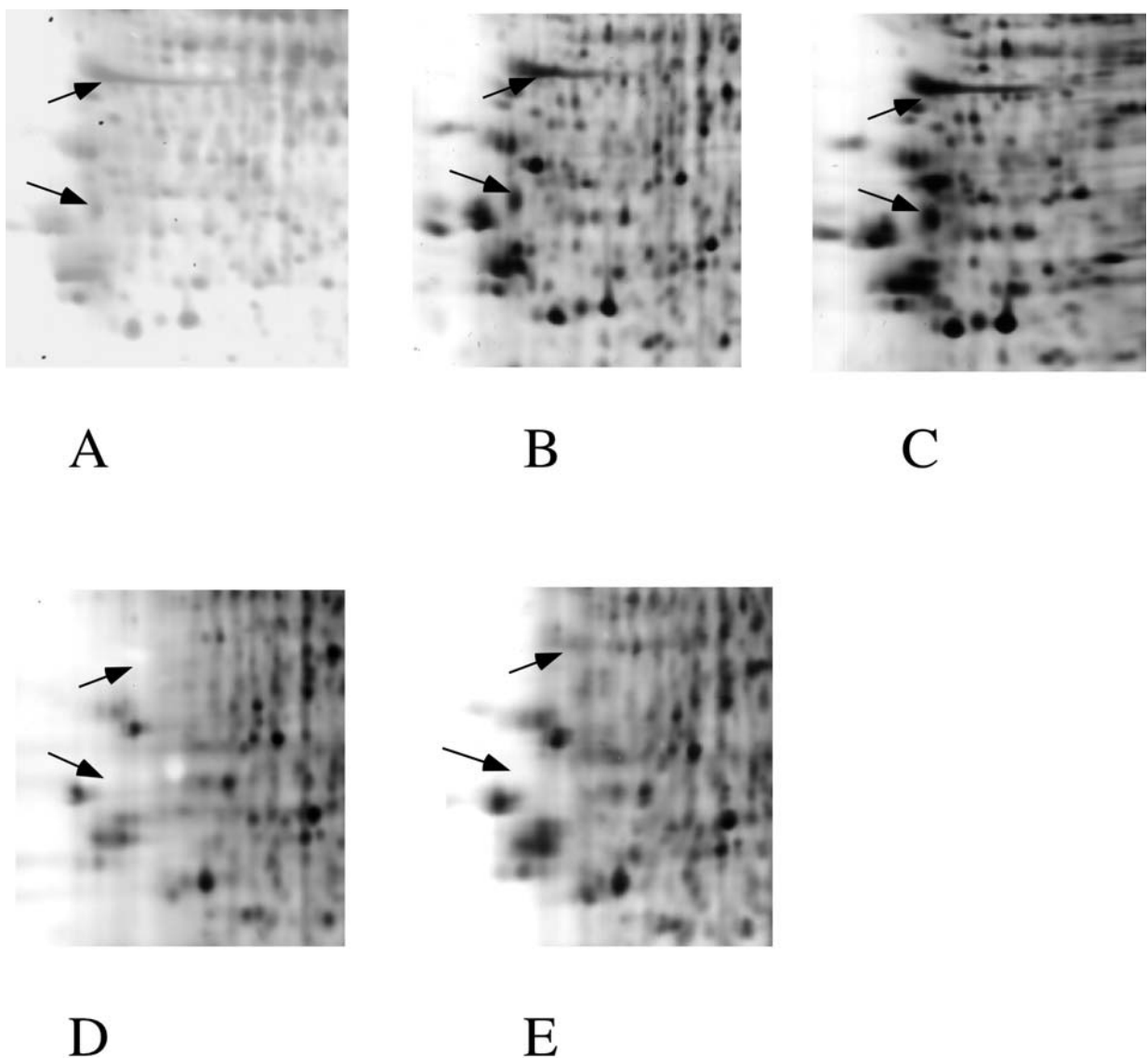

Figure 2: Detail of silver staining in the acidic region.
A small homologous gel zone is compared from the gels shown in figure 1. The gel panels are in the same order than the total gels shown on figure 1. Hollow spots (arrows) are obvious in gels stained with ammoniacal silver (panels D and E), but also in the gel stained with the ultrafast method (panel A).

15